\documentclass{elsart}
\usepackage[]{amsmath,amssymb}
\usepackage[]{epsfig}
\begin{document}
\begin{frontmatter}
\begin{flushright}
\texttt{IZTECH-P-11/02}
\end{flushright}%
\title{Stress-Energy Connection and Cosmological Constant Problem}
\author{Durmu{\c s} A. Demir}
\address{Department of Physics, {\.I}zmir Institute of Technology, TR35430, {\.I}zmir, Turkey}
 \begin{abstract}
We study gravitational properties of vacuum energy by erecting a geometry on the stress-energy tensor of vacuum, matter and radiation. Postulating that the gravitational effects of matter and radiation can be formulated by an appropriate modification of the spacetime connection, we obtain varied geometro-dynamical equations which properly comprise the usual gravitational field equations with, however, Planck-suppressed, non-local, higher-dimensional additional  terms.  The prime novelty brought about by the formalism is that, the vacuum energy does act not as the cosmological constant but as the source of the gravitational constant.  The formalism thus deafens the cosmological constant problem by channeling vacuum energy to gravitational constant. Nevertheless,   quantum gravitational effects, if any, restore the problem via the graviton and graviton-matter loops, and the mechanism proposed here falls short of taming such contributions to cosmological constant. 

\end{abstract}
\end{frontmatter}


\section{Introduction}

In regions of spacetime devoid of energy, momentum, stress or pressure distribution,
curving of the spacetime fabric is governed by the matter-free gravitational field equations
\begin{eqnarray}
\label{field0}
G_{\alpha\beta}\left(\between_V, V\right) = V_{\alpha\beta}
\end{eqnarray}
written purposefully in a slightly different form by utilizing the `metric tensor'
\begin{eqnarray}
V_{\alpha\beta} = - \Lambda_0 g_{\alpha\beta}\,.
\end{eqnarray}
This is nothing but the empty space stress-energy tensor, wherein $g_{\alpha\beta}$ is the true metric tensor on the manifold, and 
$\Lambda_0$ -- Einstein's cosmological constant (CC) \cite{einstein1} --  is the intrinsic curvature of  spacetime.  

The stress tensor of nothingness generates the connection
 \begin{eqnarray}
\label{levi-civita}
 \left(\between_V\right)^{\lambda}_{\alpha \beta} &=&  \frac{1}{2} \left( V^{-1} \right)^{\lambda\mu} \left(\partial_{\alpha} V_{\beta \mu}
+ \partial_{\beta} V_{\mu \alpha} - \partial_{\mu} V_{\alpha \beta}\right)
\end{eqnarray}
which is identical to  the Levi-Civita connection
\begin{eqnarray}
\Gamma^{\lambda}_{\alpha\beta} &=& \frac{1}{2} g^{\lambda\mu} \left(\partial_{\alpha} g_{\beta \mu}
+ \partial_{\beta} g_{\mu \alpha} - \partial_{\mu} g_{\alpha \beta}\right)
\end{eqnarray}
of the metric tensor $g_{\alpha\beta}$. This equivalence between $\between_V$ (The symbol $\between$, the letter {\it b} in Turkic runic, is short-hand for {\it bagh}  meaning `connection' in Turkish.) and $\Gamma$ holds for any value of $\Lambda_0$ provided that it is strictly constant. The connection $\between_V$ generates the Einstein tensor
\begin{eqnarray}
G_{\alpha\beta}\left(\between_V, V\right) = R_{\alpha\beta}\left(\between_V\right) - \frac{1}{2} V_{\alpha\beta} R\left(\between_V, V\right)\;\;\;\;
\end{eqnarray}
which identically equals the Einstein tensor  $G_{\alpha\beta}\left(\Gamma, g\right)$ in General Relativity (GR). In here,  $R\left(\between, V\right) \equiv  \left(V^{-1}\right)^{\mu\nu} R_{\mu\nu}\left(\between\right)$ is the Ricci scalar, $R_{\alpha\beta}\left(\between\right) \equiv R^{\mu}_{\alpha \mu \beta}\left(\between\right)$ is the Ricci tensor, and 
\begin{eqnarray}
\label{Riemann}
R^{\mu}_{\alpha\nu\beta}\left(\between\right) = \partial_{\nu} \between^{\mu}_{\beta\alpha} -  \partial_{\beta} \between^{\mu}_{\nu\alpha} + \between^{\mu}_{\nu\lambda}\between^{\lambda}_{\beta\alpha} - \between^{\mu}_{\beta\lambda}\between^{\lambda}_{\nu\alpha}\;\;\; 
\end{eqnarray}
is the Riemann tensor as generated by the connection $\between^{\lambda}_{\alpha\beta}$.

If the region of spacetime under concern is endowed with an energy, momentum, stress or pressure distribution, which
are collectively encapsulated in the stress-energy tensor $T_{\alpha\beta}$, the matter-free gravitational field equations (\ref{field0}) change to
\begin{eqnarray}
\label{field1}
G_{\alpha \beta}\left(\between_V, V\right) = V_{\alpha \beta} + 8 \pi G_N T_{\alpha \beta}
\end{eqnarray}
wherein the two sources are seen to directly add up \cite{einstein2}. In general, $T_{\alpha \beta}$ involves all the matter and force fields as well as the metric tensor $V_{\alpha\beta}$. Indeed, $T_{\alpha\beta}$ is computed from the quantum effective action which
encodes quantum fluctuations of entire matter and all forces but gravity in the background geometry determined by $V_{\alpha\beta}$. Quantum theoretic structure ensures that
\begin{eqnarray}
\label{en-mom}
T_{\alpha \beta} = - \texttt{E}\, g_{\alpha \beta} + {\texttt{t}}_{\alpha \beta}
\end{eqnarray}
where $\texttt{E}$ is the energy density of the vacuum, and ${\texttt{t}}_{\alpha \beta}$ is the
stress-energy tensor of everything but the vacuum. Putting $T_{\alpha \beta}$ into Eq. (\ref{field1}) gives rise
to an effective CC
\begin{eqnarray}
\label{lameff}
\Lambda_{\texttt{eff}} = \Lambda_0 + 8 \pi G_N \texttt{E}
\end{eqnarray}
which must nearly saturate the present expansion rate of the Universe
\begin{eqnarray}
\label{cond}
\Lambda_{\texttt{eff}} \lesssim H_0^2
\end{eqnarray}
where $H_0 \simeq 73.2\ {\rm Mpc}^{-1}\, {\rm s}^{-1}\, {\rm km}$ according to the WMAP seven-year mean \cite{astro8}.

If $\Lambda_0$ not $\Lambda_{\texttt{eff}}$ were used, the bound
(\ref{cond}) would furnish, through the observational value of
$H_0$ quoted above, an empirical determination of $\Lambda_0$,
as for any other fundamental constant of Nature. The same does
not apply to $\Lambda_{\texttt{eff}}$, however. This is because  
the vacuum energy density $\texttt{E}$, equaling the
zero-point energies of quantum fields plus enthalpy released by
various phase transitions, is much larger than
$\Lambda_{\texttt{eff}}^{\texttt{exp}}/ 8 \pi G_N$. Therefore, previously determined, 
experimentally confirmed matter and forces down to the
terascale $M_{W} \sim {\rm TeV}$, are expected to induce a vacuum energy density
of order $M_{W}^4$.  This is enormous  compared to
$\Lambda_{\texttt{eff}}^{\texttt{exp}}/ 8 \pi G_N$, and hence,
enforcement of $\Lambda_{\texttt{eff}}$ to respect the bound (\ref{cond}) introduces
a severe tuning of $\Lambda_0$ and $ 8 \pi G_N \texttt{E}$
up to at least sixty decimal places. This immense tuning becomes incrementally 
worse  as the electroweak theory is extended to higher and higher energies. As a
result, we face the biggest naturalness problem -- the cosmological constant problem (CCP) --
which plagues both particle physics and cosmology.

Over the decades, since its first solidification in \cite{ccp1},
the CCP has been approached by various proposals
and interpretations, as reviewed and critically discussed in
\cite{nobbenhuis1,review}. Each proposal involves a
certain degree of speculation in regard to going beyond (\ref{field1})
by postulating novel symmetry arguments, relaxation mechanisms, modified
gravitational dynamics and statistical interpretations \cite{nobbenhuis1,review}.
Except for the nonlocal, acausal modification of gravity implemented in \cite{nima1}
and the anthropic approach \cite{anthro}, most of the solutions proposed
for the CCP seem to overlook the already-existing vacuum energy density
${\mathcal{O}}\left[{\rm TeV}^{4}\right]$ induced by known physics down to the
terascale \cite{weinberg1}. However, any resolution of the CCP,
irrespective of how speculative it might be, must, in the first
place, provide an understanding of how this existing energy component is to be tamed.

Crystallization of the problem, as it arises in GR through (\ref{field1}),
may be interpreted to show that, {\it the CCP is actually the problem of finding the correct method for
incorporating the stress-energy tensor $T_{\alpha\beta}$ into the matter-free gravitational field equations
 so that the vacuum energy} $\texttt{E}$, {\it however large it might be, does not  contribute
to the effective CC.} Indeed, depending on how this incorporation is made, the gravitational field equations
can admit variant interpretations and maneuvers for the vacuum energy, and it might then be possible to achieve a 
resolution for the CCP.

Thus, inspired by the recent work  \cite{demirz}, {\it the present work will put forward a
novel approach to the CCP in which $T_{\alpha\beta}$ is incorporated into (\ref{field0})
by modifying not the metric $V_{\alpha\beta}$ but the connection $\left(\between_{V}\right)^{\lambda}_{\alpha\beta}$.}
Given in Sec. II below is a detailed discussion of the method.  The  unexampled concept of `stress-energy connection' 
will be introduced therein. Sec. III discusses certain questions concerning the workings of the mechanism. 
Sec. IV concludes the work.

\section{Stress-Energy Connection and Cosmological Constant}

In search for an alternative method, certain clues are provided by the scaling properties  of gravitational field equations.
Under a rigid Weyl rescaling \cite{weyl}
\begin{eqnarray}
\label{scale}
g_{\alpha \beta} \rightarrow a^2 g_{\alpha\beta}
\end{eqnarray}
the gravitational field equations (\ref{field1}) take the form
\begin{eqnarray}
\label{field1p}
G_{\alpha \beta}\left(\between_V, V\right) = a^2 V_{\alpha \beta} + 8 \pi \left(G_N a^{-2}\right) T_{\alpha \beta} \left(a^{d} \mu_{(d)}\right)
\end{eqnarray}
where $\mu_{(d)}$ is a mass dimension-$d$ coupling in the matter sector. The geometrodynamical 
variables $\left(\between_V\right)^{\lambda}_{\alpha\beta}$ and $G_{\alpha \beta}\left(\between_V, V\right)$ are strictly 
invariant under  the global rescaling (\ref{scale}). However, sources $V_{\alpha\beta}$ and 
$G_N T_{\alpha\beta}$, containing fixed scales corresponding to masses, dimensionful couplings and renormalization
scale, do  not exhibit any invariance as such.  Notably, however, even if the bare CC $\Lambda_0$ vanishes completely or 
if the matter sector possesses exact scale invariance ( $T_{\alpha \beta} \rightarrow a^{-2} T_{\alpha\beta}$),
gravitational field equations are  never Weyl invariant  simply because Newton's constant is there to scale 
as $a^{-2}$. 

A short glance at (\ref{field1p}) reveals that, the combination 
\begin{eqnarray}
\label{field1px}
G_{\alpha \beta}\left(\between_V, V\right) - 8 \pi \left(G_N a^{-2}\right) T_{\alpha \beta} \left(a^{d} \mu_{(d)}\right)\;\;\;
\end{eqnarray}
owns the transformation property of  the Einstein tensor pertaining to a non-Riemannian geometry. This is readily seen
by noting that, a general connection $\between$ can always decomposed as 
\begin{eqnarray}
\label{eqn3}
\between^{\lambda}_{\alpha\beta} = \left(\between_V\right)^{\lambda}_{\alpha\beta} + \Delta^{\lambda}_{\alpha\beta}
\end{eqnarray}
where $ \Delta$  is a rank (1,2) tensor field. In response to this split structure, the Einstein tensor of $\between$ breaks up into two
\begin{eqnarray}
\label{einsteinx}
{\mathbb{G}}_{\alpha\beta}\left(\between, V\right) = G_{\alpha \beta}\left(\between_V, V\right) + {\mathcal{G}}_{\alpha \beta}\left(\Delta, V\right)\;\;\;
\end{eqnarray}
where ${\mathcal{G}}_{\alpha\beta}\left(\Delta, V\right) $, not found in GR, reads as
\begin{eqnarray}
\label{einstein-extra}
{\mathcal{G}}_{\alpha\beta}\left(\Delta, V\right) = {\mathcal{R}}_{\alpha\beta}(\Delta) - \frac{1}{2} V_{\alpha\beta} \left(V^{-1}\right)^{\mu \nu} {\mathcal{R}}_{\mu\nu}(\Delta)
\end{eqnarray}
with
\begin{eqnarray}
\label{ricci-extra}
{\mathcal{R}}_{\alpha \beta}\left(\Delta\right) = \nabla_{\mu}
\Delta^{\mu}_{\alpha \beta} - \nabla_{\beta} \Delta^{\mu}_{\mu
\alpha} + \Delta^{\mu}_{\mu
\nu} \Delta^{\nu}_{\alpha \beta} - \Delta^{\mu}_{\beta
\nu} \Delta^{\nu}_{\alpha \mu}\,.
\end{eqnarray}
Under the global scaling in (\ref{scale}), $G_{\alpha \beta}\left(\between_V, V\right)$ stays at its original value yet ${\mathcal{G}}_{\alpha \beta}\left(\Delta, V\right)$ exhibits modifications contingent on how $\Delta^{\lambda}_{\alpha\beta}$ depends on the metric tensor. Formally,  the Einstein tensor in (\ref{einsteinx}) changes to  
\begin{eqnarray}
\label{field1pp}
G_{\alpha \beta}\left(\between_V, V\right) + {\mathcal{G}}_{\alpha \beta}\left(\Delta(a), V\right)
\end{eqnarray} 
which obtains the same structure as the combination in (\ref{field1px}) as far as the scaling properties
of individual terms are concerned. 

At this point,  there arises a crucial question as to whether their formal similarity under scaling can ever promote  (\ref{field1pp}) to a novel formulation alternative to (\ref{field1px}). In other words, can part of (\ref{field1px}) involving  the stress-energy tensor arise, partly or wholly, from ${\mathcal{G}}_{\alpha \beta}\left(\Delta, V\right)$ ? Can matter and radiation be put in interaction with gravity by enveloping $T_{\alpha\beta}$ into connection instead of  adding it to $V_{\alpha\beta}$ as in (\ref{field1}) ? These questions, which are at the heart of the novel formulation being constructed,  cannot be answered without a proper understanding of the tensorial connection $\Delta$. To this end, one observes that generating $T_{\alpha\beta}$ from 
${\mathcal{G}}_{\alpha \beta}\left(\Delta, V\right)$ can be a quite intricate process since while $T_{\alpha\beta}$ is divergence-free  ${\mathcal{G}}_{\alpha \beta}\left(\Delta, V\right)$ is not 
\begin{eqnarray}
\label{non-conserve}
\nabla^{\alpha} {\mathcal{G}}_{\alpha \beta}\left(\Delta, V\right) \neq 0
\end{eqnarray}
because $ {\mathcal{R}}_{\alpha \beta}\left(\Delta\right)$, as it is not generated by commutators of $\nabla^{\between_V}$ or $\nabla^{\between}$,  is not necessarily a true curvature tensor to obey the Bianchi identities. For relating $\Delta^{\lambda}_{\alpha\beta}$ to  $T_{\alpha\beta}$, it proves facilitative to  introduce a symmetric tensor field 
\begin{eqnarray}
\label{mat-T}
\mathbb{T}_{\alpha\beta} = - \Lambda g_{\alpha\beta} + \Theta_{\alpha \beta}
\end{eqnarray}
which will be related to $T_{\alpha\beta}$ in the sequel. For definiteness, $\mathbb{T}_{\alpha\beta}$, similar to the stress-energy tensor $T_{\alpha\beta}$,  is split into a covariantly-constant part which is its first term ($\Lambda$ is strictly constant), and a generic symmetric tensor field $\Theta_{\alpha\beta}$ which does, by construction, not contain any  covariantly-constant structure.  As an obvious  way of incorporating $T_{\alpha\beta}$ into (\ref{field0}) via $\between$, one can write
\begin{eqnarray}
\label{new-conn}
\between^{\lambda}_{\alpha\beta} = \left(\between_{V + \mathbb{T}}\right)^{\lambda}_{\alpha\beta}
\end{eqnarray}
which follows from (\ref{levi-civita}) by replacing $V_{\alpha\beta}$ therein with $V_{\alpha\beta} + {\mathbb{T}}_{\alpha\beta}$. As a result, $\Delta$   becomes
\begin{eqnarray}
\label{tensor-conn}
\Delta^{\lambda}_{\alpha \beta} &=& \frac{1}{2} {\left(\left(V+{\mathbb{T}}\right)^{-1}\right)}^{\lambda \nu}
 \left( \nabla_{\alpha} {\mathbb{T}}_{\beta\nu} + \nabla_{\beta} {\mathbb{T}}_{\nu \alpha} - \nabla_{\nu} {\mathbb{T}}_{\alpha \beta}\right)\nonumber\\
&=&  \frac{1}{2} {\left(\left(V+{\mathbb{T}}\right)^{-1}\right)}^{\lambda \nu} \left( \nabla_{\alpha} \Theta_{\beta\nu} + \nabla_{\beta} \Theta_{\nu \alpha} - \nabla_{\nu} \Theta_{\alpha \beta}\right)
\end{eqnarray}
which is seen to be a sensitive probe of  $\Theta_{\alpha\beta}$ since it vanishes identically as $\Theta_{\alpha\beta} \rightarrow 0$. 
 
By way of (\ref{new-conn}), the Einstein tensor in (\ref{field1pp}) takes the form
\begin{eqnarray}
\label{field1ppp}
G_{\alpha \beta}\left(\between_V, V\right) + {\mathcal{G}}_{\alpha \beta}\left((\Lambda_0 + \Lambda) a^2, \Theta(a)\right)\;\;\;
\end{eqnarray}
whose comparison with (\ref{field1p}) reveals the following features:  
\begin{enumerate}
\item The parameter $\Lambda$ in (\ref{mat-T}) must be related to the gravitational constant $G_N$. Actually, a relation of the form
\begin{eqnarray}
\label{newton}
\Lambda + \Lambda_0 = (8 \pi G_N)^{-1}
\end{eqnarray}
 is expected on general grounds. 

\item  In the limit $T_{\alpha\beta} \rightarrow 0$, the gravitational field equations (\ref{field1})  uniquely reduce to the matter-free field equations (\ref{field0}). Likewise, the gravitational field equations to be obtained here, as suggested by (\ref{new-conn}), must smoothly reduce to (\ref{field0}) as $\mathbb{T} \rightarrow 0$. Therefore, any functional relation $\mathbb{T}_{\alpha\beta}= \mathbb{T}_{\alpha\beta}[T]$ between $\mathbb{T}$ and $T$ should exhibit the correspondence 
\begin{eqnarray}
\label{cond-zero}
 T_{\alpha\beta} = 0 \Longleftrightarrow \mathbb{T}_{\alpha\beta} = 0\,.
\end{eqnarray}
In addition, as $T_{\alpha\beta} \rightarrow - \texttt{E} g_{\alpha\beta}$, the right-hand side of (\ref{field0}) changes to $\left(1+ \texttt{E}/\Lambda_0\right) V_{\alpha\beta}$, which clearly signals the CCP. In contrast, however, as $\mathbb{T}_{\alpha\beta} \rightarrow - \Lambda g_{\alpha\beta}$, 
 $\mathbb{G}_{\alpha\beta}\left( \between_{V + {\mathbb{T}}}, {V}\right)$ reduces to the matter-free form $\mathbb{G}_{\alpha\beta}\left( \between_V, V\right)$. In other words, even if matter and radiation are discarded, that is, $T_{\alpha\beta} = - \Lambda g_{\alpha\beta}$ ($\texttt{t}_{\alpha\beta} = 0$),  the gravitational field equations (\ref{field1}) suffer from the CCP.  However, when $\mathbb{T}_{\alpha\beta} = - \Lambda g_{\alpha\beta}$ ($\Theta_{\alpha\beta} = 0$), $ \mathbb{G}_{\alpha\beta}\left( \between_{V + {\mathbb{T}}}, {V}\right)$ remains unchanged at  $\mathbb{G}_{\alpha\beta}\left( \between_V, V\right)$ with complete immunity to $\Lambda$. 
\end{enumerate}
These observations evidently reveal the physical and CCP-wise relevance of the method. 

As a matter of course, the  dynamical equation
\begin{eqnarray}
\label{eom-new}
{\mathbb{G}}_{\alpha\beta}\left(\between_{V+{\mathbb{T}}}, V\right) = V_{\alpha\beta}\,,
\end{eqnarray}
as directly follows from  (\ref{field0}) via the replacement $\between_V \rightarrow \between_{V + {\mathbb{T}}}$, forms the germ of the CCP-free gravitational dynamics under attempt. Under (\ref{einsteinx}),  it gives 
\begin{eqnarray}
\label{eom}
G_{\alpha \beta}\left(\between_V, V\right) = V_{\alpha\beta} - {\mathcal{G}}_{\alpha\beta}\left(\Delta, V\right)
\end{eqnarray}
which refines the germinal equation (\ref{eom-new}). To proceed further, it is necessary to establish the relation between $\mathbb{T}_{\alpha\beta}$ and $T_{\alpha\beta}$ so that (\ref{eom-new}) reduces to (\ref{field1}), at least approximately. This reduction does of course not affect the value of CC; it stays put at $\Lambda_0$.  On the other hand, with (\ref{newton}) relating $\Lambda$ to $G_N$, on physical grounds, one expects $\left|\Lambda\right| \gg \left|\Theta\right|$. Then,  all quantities, in particular,  $\Delta^{\lambda}_{\alpha\beta}$ can be expanded in powers of   $\Theta/\Lambda$ such that (\ref{eom}), at the leading order, is to return the gravitational field equations (\ref{field1}). As a matter of fact, the dynamical equation (\ref{eom}), after using
\begin{eqnarray}
{\left(\left(V+{\mathbb{T}}\right)^{-1}\right)}_{\alpha\beta} &=&  (8\pi G_N) g_{\alpha\beta} - (8 \pi G_N)^2 \Theta_{\alpha\beta}
+ (8\pi G_N)^3 \Theta^{\mu}_{\alpha} \Theta_{\mu \beta} - \dots
\end{eqnarray}
takes the form
\begin{eqnarray}
\label{eompp}
G_{\alpha \beta}\left(\between_V, V\right) &=& \texttt{C}^{(0)}_{\alpha\beta} + (8\pi G_N) \texttt{C}^{(1)}_{\alpha\beta} + (8\pi G_N)^2 \texttt{C}^{(2)}_{\alpha\beta} + \dots
\end{eqnarray}
where $\texttt{C}^{(n)}_{\alpha\beta}$ are valency-two symmetric tensor fields encapsulating all the terms of order $(8\pi G_N)^n$.  For $n=0$, the tensorial connection $\Delta^{\lambda}_{\alpha\beta}$ vanishes identically, and hence,
\begin{eqnarray}
\label{C0value}
\texttt{C}^{(0)}_{\alpha\beta} = V_{\alpha\beta} 
\end{eqnarray}
so that (\ref{eom-new}) directly reduces to the matter-free gravitational field equations (\ref{field0}) for $\mathbb{T}_{\alpha\beta} = 0$
as well as $\mathbb{T}_{\alpha\beta} = - \Lambda g_{\alpha\beta}$.

For $n=1$,
\begin{eqnarray}
\Delta^{\lambda}_{\alpha\beta} = 4 \pi G_N ( \nabla_{\alpha} {\Theta}^{\lambda}_{\beta} + \nabla_{\beta} {\Theta}^{\lambda}_{\alpha} - \nabla^{\lambda}
{\Theta}_{\alpha \beta})\;\;\;\;
\end{eqnarray}
 is linear in ${\Theta}_{\alpha \beta}$,  and so is the derivative part of ${\cal{R}}_{\alpha \beta}\left(\Delta\right)$. Then, ${\mathcal{G}}_{\alpha\beta}\left(\Delta, V\right)$ defined in (\ref{einstein-extra}) yields
\begin{eqnarray}
\label{C1value}
\texttt{C}^{(1)}_{\alpha\beta} = - 2 \left[\texttt{K}^{-1}\left(\nabla\right)\right]_{\alpha\beta}^{\mu\nu} {\Theta}_{\mu \nu}
\end{eqnarray}
where
\begin{eqnarray}
\label{inv-prop}
\left[\texttt{K}^{-1}\right]_{\alpha\beta\mu\nu}(\nabla) &=& \frac{1}{8}\left(\nabla_{\mu}\nabla_{\alpha} g_{\nu\beta}+  \nabla_{\mu}\nabla_{\beta} g_{\alpha\nu}\right) +
 \frac{1}{8}\left(\nabla_{\nu} \nabla_{\alpha} g_{\mu\beta} + \nabla_{\nu}\nabla_{\beta} g_{\alpha\mu}\right)\nonumber\\
&-&\frac{1}{8}\left( \nabla_{\alpha}\nabla_{\beta} + \nabla_{\beta}\nabla_{\alpha}\right) g_{\mu\nu} -
\frac{1}{8}  \left( \nabla_{\mu}\nabla_{\nu} + \nabla_{\nu}\nabla_{\mu}\right) g_{\alpha\beta} \nonumber\\
&-&\frac{1}{8} {\Box} \left( g_{\alpha\mu}g_{\beta\nu} + g_{\alpha\nu}g_{\mu\beta} - 2 g_{\alpha\beta} g_{\mu\nu}\right) 
\end{eqnarray}
is nothing but the inverse propagator for a `massless spin-2 field' in the background geometry generated by $g_{\alpha\beta}$. To  reproduce the gravitational field equations (\ref{field1}) correctly, one must impose
\begin{eqnarray}
\label{ek-cond}
- 2 \left[\texttt{K}^{-1}\left(\nabla\right)\right]_{\alpha\beta}^{\mu\nu} {\Theta}_{\mu \nu} = - 2 \left[\texttt{K}^{-1}\left(\nabla\right)\right]_{\alpha\beta}^{\mu\nu} {\mathbb{T}}_{\mu \nu}  \equiv  \texttt{t}_{\alpha\beta}
\end{eqnarray}
where $`` \texttt{t}_{\alpha\beta} "$ was defined in  (\ref{en-mom}) to involve `no covariantly-constant part'. This equality lies at the
heart of the mechanism being proposed, and therefore, its analysis and examination prove vital for further progress.  The main question is this: Can the right-hand side of (\ref{ek-cond}) ever involve a covariantly-constant part (of the form $c_1 g_{\alpha\beta}$ with $c_1$ constant)  added to $\texttt{t}_{\alpha\beta}$? If the answer turns out to be affirmative then whole mechanism collapses down since $c_1/M^2$, unless guaranteed to lie near $\Lambda_0$ by some reason, brings back the CCP. In examining, one first notes that the equality (\ref{ek-cond}) works fine for both $ {\Theta}_{\alpha \beta}$ and $ {\mathbb{T}}_{\alpha \beta}$ since a covariantly-constant part (like $\Lambda g_{\alpha\beta}$) is automatically nullified by $\left[\texttt{K}^{-1}\left(\nabla\right)\right]_{\alpha\beta}^{\mu\nu}$. Therefore, if $\texttt{t}_{\alpha\beta}$ in (\ref{ek-cond}) is to change to  $\texttt{t}_{\alpha\beta} + c_1 g_{\alpha\beta}$ there has to be an appropriate structure within ${\mathbb{T}}_{\alpha \beta}$. The requisite structure is found to be 
\begin{eqnarray}
\delta {\mathbb{T}}_{\alpha \beta} = \left[\texttt{K}\left(\nabla\right)\right]_{\alpha\beta}^{\mu\nu}\left(c_1 g_{\mu \nu}\right) \equiv k_1 g_{\alpha\beta}
\end{eqnarray}
where structure of the spin-2 propagator $\left[\texttt{K}\left(\nabla\right)\right]_{\alpha\beta}^{\mu\nu}$ guarantees that  $k_1 = \pm \infty$ independent of the value of $c_1$. This result implies that the covariantly-constant part of ${\mathbb{T}}_{\alpha \beta}$ in (\ref{mat-T}) changes to $(\Lambda + k ) g_{\alpha\beta}\equiv \Lambda_{eff} g_{\alpha\beta}$ with $\Lambda_{eff} = \pm  \infty$. In other words, an infinite $\Lambda$ in  ${\mathbb{T}}_{\alpha \beta}$ corresponds to a covariantly-constant part of the form $c_1 g_{\alpha\beta}$ in (\ref{ek-cond}). However, $\Lambda \rightarrow \pm \infty$ in ${\mathbb{T}}_{\alpha \beta}$ causes the tensorial connection $\Delta^{\lambda}_{\alpha\beta}$ in (\ref{tensor-conn}) to vanish, and hence,  the germinal equation (\ref{eom-new}) to reduce to the original matter-free gravitational field equations (\ref{field0}). This implies that an infinite $\Lambda$ prohibits the incorporation of matter and radiation into (\ref{field0}). These observations and findings should provide enough evidence that  $`` \texttt{t}_{\alpha\beta} "$ in (\ref{ek-cond}) is the stress-energy tensor of everything but vacuum; it cannot have a covariantly-constant part. It is precisely what was meant in writing (\ref{mat-T}), and hence, everything but vacuum gravitates precisely as in the GR. Obviously, ${\Theta}_{\alpha \beta}$ is related to $\texttt{t}_{\alpha\beta}$ non-locally yet causally since ${\Theta}_{\alpha \beta}$ involves values of $\texttt{t}_{\alpha\beta}$ in every place and time as propagated by the `massless spin-2 propagator' $\texttt{K}_{\alpha\beta\mu\nu}\left(\nabla\right)$. By inverting (\ref{ek-cond}) one finds
\begin{eqnarray}
\mathbb{T}_{\alpha\beta} = \Theta^{0}_{\alpha\beta} - \frac{1}{2} \left[\texttt{K}\left(\nabla\right)\right]_{\alpha\beta}^{\mu\nu} \texttt{t}_{\mu\nu}
\end{eqnarray}
where $\Theta^{0}_{\alpha\beta}\equiv - \Lambda g_{\alpha\beta}$ is covariantly-constant. In fact, it must be proportional to the vacuum energy density in (\ref{en-mom}), that is, $\Theta^{0}_{\alpha\beta} \propto \texttt{E} g_{\alpha\beta}$. Consequently,
\begin{eqnarray}
\label{soln-mathbbT}
\mathbb{T}_{\alpha\beta} = - {\texttt{L}}^2 \texttt{E} g_{\alpha \beta} - \frac{1}{2} \left[\texttt{K}\left(\nabla\right)\right]_{\alpha\beta}^{\mu\nu} \texttt{t}_{\mu\nu}
\end{eqnarray}
wherein $\Lambda =  {\texttt{L}}^2 \texttt{E}$, and $\texttt{L}^2$, having the dimension of area, arises for dimensionality reasons. This expression establishes a direct relationship between  $\mathbb{T}_{\alpha\beta}$ and $T_{\alpha\beta}$ so that  $\mathbb{T}_{\alpha\beta} = 0 \Longleftrightarrow T_{\alpha\beta} = 0$, as was discussed in detail in relation to (\ref{cond-zero}). Actually, it is possible
to interpret the result (\ref{soln-mathbbT}) in a more general setting by generalizing the propagator (\ref{inv-prop}) to massive case
\begin{eqnarray}
\label{massive-inv-prop}
\left[{\mathcal{K}}^{-1}\right]_{\alpha\beta\mu\nu} \left(\nabla, {\texttt{L}}^2\right) &=&  \left[\texttt{K}^{-1}\right]_{\alpha\beta\mu\nu}(\nabla)\nonumber\\
&-&\frac{f\left({\texttt{L}}^2 \Box\right)}{4 {\texttt{L}}^2} \left(g_{\alpha\beta} g_{\mu\nu} - g_{\alpha\mu}g_{\beta\nu} - g_{\alpha\nu}g_{\mu\beta}\right)
\end{eqnarray}
where the operator $f\left({\texttt{L}}^2 \Box\right)/{\texttt{L}}^2$ serves as the `mass-squared' parameter with the distributional structure
\begin{eqnarray}
f(x)=\left\{\begin{array}{c} 1,\ {x}=0\\ 0,\ {x\neq 0}\end{array}\right.
\end{eqnarray}
similar to the one used in \cite{nima1}. In (\ref{massive-inv-prop}), care is needed in interpreting the `mass term' in that there is actually no `spin-2 mass term' to speak about:  It vanishes for non-uniform sources like $\texttt{t}_{\alpha\beta}$ and stays constant for uniform sources like $\Lambda g_{\alpha\beta}$.  Clearly,  the `massive propagator' above automatically reproduces the result in (\ref{soln-mathbbT})
\begin{eqnarray}
\label{f-rel}
\mathbb{T}_{\alpha\beta} &=& \left[{\mathcal{K}}\right]_{\alpha\beta}^{\mu\nu} \left(\nabla, {\texttt{L}}^2\right) T_{\mu \nu}
= - {\texttt{L}}^2 \texttt{E} g_{\alpha\beta} - ({1}/{2}) \texttt{K}\left(\nabla\right)_{\alpha\beta}^{\mu\nu} \texttt{t}_{\mu \nu}
\end{eqnarray}
thanks to  the distributional structure of $f(x)$.

For $n=2$ and higher, the tensorial connection $\Delta^{\lambda}_{\alpha\beta}$ goes like $\Theta^{n-1}$ times $\nabla \Theta$, and is always proportional to $\Delta (n=1)$. More explicitly,
\begin{eqnarray}
\label{tensor-conn-n}
\Delta^{\lambda}_{\alpha\beta}(n) = \left[ \Pi_{k=1}^{n-1} (-8  \pi G_N)^{k} \Theta^{\lambda}_{\mu_{k}}\right]  \Delta^{\mu_{1}}_{\alpha\beta} \left(1\right)
\end{eqnarray}
where each $\Theta$ factor is expressed in terms of $\texttt{t}$ via (\ref{f-rel}).  Gradients of $\Delta^{\lambda}_{\alpha\beta}(n)$ and bilinears  $\left[\Delta(n-k) \otimes \Delta(k)\right]_{\alpha\beta}$ ($k=1,2,\dots,n-1$) add up to form $\texttt{C}^{(n)}_{\alpha\beta}$ in accord with the structure of ${\mathcal{G}}_{\alpha\beta}\left(\Delta\right)$ in (\ref{einstein-extra}).  In contrast to the three tensor fields $G_{\alpha \beta}(\between_V, V)$, $\texttt{C}^{(0)}_{\alpha\beta}$  and $\texttt{C}^{(1)}_{\alpha\beta}$,  it is not clear if  $\texttt{C}^{(n\geq 2)}_{\alpha\beta}$ acquires vanishing divergence, in general.  Therefore, the gravitational field equations 
\begin{eqnarray}
\label{eomppp}
G_{\alpha \beta}= - \Lambda_0 g_{\alpha\beta} + (8\pi G_N) \texttt{t}_{\alpha\beta} + {\mathcal{O}}\left[\left(8\pi G_N \nabla\Theta\right)^2, \left(8\pi G_N\right)^2 \Theta \nabla \nabla\Theta\right]
\end{eqnarray}
distilled from the germinal dynamics in (\ref{eom-new}), are insensitive to vacuum energy density $\texttt{E}$ yet suffer from a serious inconsistency that the divergence of $\texttt{C}^{(n\geq 2)}_{\alpha\beta}$ may not vanish at all. The next section will give a critique of the formalism, as developed so far.

\section{More on the Formalism}

Comparison of (\ref{eomppp})  with (\ref{field1}) raises certain questions pertaining to the consistency of the elicited gravitational dynamics. There are mainly three questions:

{\bf Question 1.} What precludes ${\mathcal{G}}_{\alpha\beta}\left(\Delta, V\right)$ from developing a covariantly-constant part that can act as the CC?

{\bf Question 2.} What must be the structure of $\mathbb{T}_{\alpha\beta}$ such that, despite Eq.(\ref{non-conserve}), $\nabla^{\alpha} {{G}}_{\alpha \beta}\left(\Delta, V\right) $ is nullified to make both sides of (\ref{eom}) divergence-free?

{\bf Question 3.} What is the status of CCP under the formalism developed here? 

Answers to these questions will disclose the physical meaning, scope and reach of the gravitational field equations (\ref{eomppp}).

\subsection{ Answer to Question 1}
 It is of prime importance to determine if the quasi Einstein tensor ${\mathcal{G}}_{\alpha\beta}\left(\Delta, V\right)$ can develop a covariantly-constant part since this type of contribution can cause the CCP.

As the definition of $\Delta^{\lambda}_{\alpha\beta}$ in (\ref{tensor-conn}) manifestly shows, $\Lambda$, in whatever way it might be related to $\texttt{E}$,   does not provide any contribution to CC. In fact, a  nontrivial $\Delta^{\lambda}_{\alpha\beta}$ originates from $\Theta_{\alpha\beta}$ only. Though it vanishes identically for  $\Theta_{\alpha\beta} = 0$, it  remains nonvanishing even for $\Lambda = 0$.  Therefore, ${\mathcal{G}}_{\alpha\beta}\left(\Delta, V\right)$ depends critically on $\Theta_{\alpha\beta}$, and any value it takes, covariantly-constant or otherwise, is governed by $\Theta_{\alpha\beta}$. There is no such sensitivity to $\Lambda$.  

As dictated by the structure of the quasi curvature tensor ${\mathcal{R}}_{\alpha\beta}$ in (\ref{ricci-extra}),  for $G_{\alpha \beta}\left(\Delta, V\right)$ to develop a covariantly-constant part,  at least one of  
\begin{eqnarray}
\label{structures}
\nabla_{\mu} \Delta^{\mu}_{\alpha \beta}\,,\; \Delta^{\mu}_{\mu \nu} \Delta^{\nu}_{\alpha \beta}\,,\;
\nabla_{\beta} \Delta^{\mu}_{\mu \alpha}\,,\; \Delta^{\mu}_{\beta \nu} \Delta^{\nu}_{\alpha \mu}\;\; 
\end{eqnarray}
must be partly proportional to the metric tensor  $g_{\alpha\beta}$ or must partly take a constant value when contracted with the metric tensor. Concerning the first and second structures above, a reasonable ansatz is $\Delta^{\lambda}_{\alpha\beta} \ni U^{\lambda} g_{\alpha\beta}$ where $U^{\alpha}$ is a vector field. With this structure for $\Delta^{\lambda}_{\alpha\beta}$, all one needs is to set $\nabla_{\mu} U^{\mu} = c_1$ for $\nabla_{\mu} \Delta^{\mu}_{\alpha \beta} \ni c_1 g_{\alpha\beta}$, and $U_{\mu} U^{\mu} = c_2$ for $\Delta^{\mu}_{\mu \nu} \Delta^{\nu}_{\alpha \beta} \ni c_2 g_{\alpha\beta}$, where $c_1$ and $c_2$ are constants. With the same ansatz for $\Delta^{\lambda}_{\alpha\beta}$, the remaining  terms in (\ref{structures}) give rise to a covariantly-constant part in ${\mathcal{G}}_{\alpha\beta}\left(\Delta, V\right)$ not by themselves but via $V_{\alpha\beta} \left(V^{-1}\right)^{\mu\nu} {\mathcal{R}}_{\mu\nu}\left(\Delta\right)$. Indeed, $\nabla_{\beta} \Delta^{\mu}_{\mu \alpha} \ni \nabla_{\beta}U_{\alpha}$ and $\Delta^{\mu}_{\beta \nu} \Delta^{\nu}_{\alpha \mu} \ni U_{\alpha}U_{\beta}$, and they contract to $c_1$ and $c_2$  for $\nabla_{\mu} U^{\mu} = c_1$ and $U_{\mu} U^{\mu} = c_2$, respectively.  A more accurate ansatz for a symmetric tensorial connection would be
\begin{eqnarray}
\label{tensor-conn-2}
{\widetilde{\Delta}}^{\lambda}_{\alpha\beta}= a U^{\lambda} g_{\alpha\beta} + b \left( \delta^{\lambda}_{\alpha} U_{\beta} + U_{\alpha} \delta^{\lambda}_{\beta}\right)\,.
\end{eqnarray}
As follows from (\ref{ricci-extra}), the Ricci tensor ${\widetilde{\mathcal{R}}}_{\alpha\beta}$ for this particular connection becomes symmetric for $a = - 5 b$,  and the Einstein tensor
\begin{eqnarray}
{\widetilde{\mathcal{G}}}_{\alpha\beta} = b \left( \nabla_{\alpha} U_{\beta} + \nabla_{\beta} U_{\alpha}\right) - 22 b^2 U_{\alpha} U_{\beta} +  b \nabla\cdot U g_{\alpha\beta} + b^2 U\cdot U g_{\alpha\beta}
\end{eqnarray}
contributes to the CC by its third term in an amount $\delta \Lambda_0 =  4 b c_1$ if $\nabla_{\mu} U^{\mu} = c_1$, and by its fourth term in an amount $\delta \Lambda_0 =  - b^2  c_2$ if $U_{\mu} U^{\mu} = c_2$. These results ensure that, at least for a connection in the form of (\ref{tensor-conn-2}),   the CCP could be resurrected  depending on how the contribution of $U^{\mu}$ compares with the bare term $\Lambda_0$.  To this end, being a symmetric tensorial connection with symmetric Ricci tensor, ${\widetilde{\Delta}}^{\lambda}_{\alpha\beta}$ in (\ref{tensor-conn-2}) can be directly compared to   $\Delta^{\lambda}_{\alpha\beta}$ in (\ref{tensor-conn}) to find 
\begin{eqnarray}
\label{trace-eq}
\frac{1}{2} \nabla_{\alpha} \log\left( \mbox{Det}\left[{\mathbb{T}}\right] \right)= \widetilde{\Delta}^{\mu}_{\mu \alpha} = 0\;\;\;\;
\end{eqnarray}
and
\begin{eqnarray}
\label{contract-eq}
\frac{1}{2} \left( {\mathbb{T}}^{-1}\right)^{\lambda \rho} \left( 2 \nabla^{\alpha} {\mathbb{T}}_{\alpha\rho} - \nabla_{\rho} {\mathbb{T}}^{\alpha}_{\alpha}\right) = g^{\alpha\beta} {\widetilde{\Delta}}^{\lambda}_{\alpha\beta} = - 18 b U^{\lambda}\,.
\end{eqnarray}
The first condition, namely the one in (\ref{trace-eq}), requires $\mathbb{T}_{\alpha\beta} = \widetilde{c}\ g_{\alpha\beta}$ where $\widetilde{c}$ is a constant. In other words, (\ref{trace-eq}) enforces $\Theta_{\alpha\beta} =0$, and its replacement in (\ref{contract-eq}) consistently gives $b =0$. Therefore, at least for connections structured like (\ref{tensor-conn-2}), there does not exist a $\Theta_{\alpha\beta}$ to equip ${\mathcal{G}}_{\alpha\beta}\left(\Delta, V\right)$ with a covariantly-constant part. 

Despite the firmness of this result, one notices that, it is actually not necessary to force $\Delta^{\lambda}_{\alpha\beta}$ to be wholly equal to ${\widetilde{\Delta}}^{\lambda}_{\alpha\beta}$ since it is sufficient to have only part of ${\mathcal{G}}_{\alpha\beta}\left(\Delta, V\right)$ be covariantly-constant. Thus, in general, one can write 
\begin{eqnarray}
\Delta^{\lambda}_{\alpha\beta} = {\widetilde{\Delta}}^{\lambda}_{\alpha\beta} + {\mathcal{D}}^{\lambda}_{\alpha\beta}
\end{eqnarray}
where ${\mathcal{D}}^{\lambda}_{\alpha\beta}={\mathcal{D}}^{\lambda}_{\beta\alpha}$, and $\nabla_{\beta} {\mathcal{D}}^{\mu}_{\mu\alpha}$ $=\nabla_{\alpha} {\mathcal{D}}^{\mu}_{\mu\beta}$ for ${\mathcal{R}}_{\alpha\beta}\left({\mathcal{D}}\right) = {\mathcal{R}}_{\beta\alpha}\left({\mathcal{D}}\right)$. This condition enforces either ${\mathcal{D}}^{\mu}_{\mu\alpha} = 0$  or ${\mathcal{D}}^{\mu}_{\mu\alpha}= \nabla_{\alpha} \Phi$, $\Phi$ being a scalar. The former, which was used for   ${\widetilde{\Delta}}^{\lambda}_{\alpha\beta}$ in (\ref{tensor-conn-2}), does not change the present conclusion. The latter,  which was used for   ${{\Delta}}^{\lambda}_{\alpha\beta}$ in (\ref{tensor-conn}), guarantees that $\Delta^{\lambda}_{\alpha\beta}$ and ${\mathcal{D}}^{\lambda}_{\alpha\beta}$ are identical up to some determinant-preserving transformations. More accurately, while ${{\Delta}}^{\lambda}_{\alpha\beta}$ makes use of $\mathbb{T}_{\alpha\beta}$, ${\mathcal{D}}^{\lambda}_{\alpha\beta}$ involves ${\mathcal{T}}_{\alpha\beta}$ which must equal  ${\mathbb{M}}_{\alpha}^{\mu} {\mathbb{T}}_{\mu \nu} \left({\mathbb{M}}^{-1}\right)^{\nu}_{\beta}$ with $\mathbb{M}_{\alpha\beta}$ being a generic tensor field. All these results ensure that, ${{\Delta}}^{\lambda}_{\alpha\beta}$ cannot cause ${\mathcal{G}}_{\alpha\beta}\left(\Delta, V\right)$ to develop a covariantly-constant part, at least for tensorial connections of the form (\ref{tensor-conn-2}).

\subsection{ Answer to Question 2}
The left-hand side of (\ref{eomppp}) is divergence-free by the Bianchi identities; however, its right-hand side exhibits no such property for $n\geq 2$. Indeed, unlike GR wherein the right-hand side obtains vanishing divergence by the conservation of matter and radiation flow,
the right-hand side of (\ref{eomppp}) lacks such a property because the quasi curvature tensor ${\cal R}^{\mu}_{\alpha\nu \beta}\left(\Delta\right)$ does not obey the Bianchi identities.  A remedy to this conservation problem, an aspect that the initiator work \cite{demirz} was lacking, comes via the expansion
\begin{eqnarray}
\label{theta-expand}
\mathbb{T}_{\alpha\beta} = - \Lambda \sum_{n=0}^{\infty} {{\left(- 8\pi G_N\right)}^{n}} {{\Theta}^{(n)}_{\alpha\beta}}
= - \Lambda g_{\alpha\beta} + {{\Theta}^{(1)}_{\alpha\beta}} -  {{\left(8\pi G_N\right)}} {{\Theta}^{(2)}_{\alpha\beta}} + \dots
\end{eqnarray}
over a set of tensor fields $\big\{{\Theta}^{(0)}_{\alpha\beta} \equiv g_{\alpha\beta},$ $\Theta^{(1)}_{\alpha\beta},$ $\Theta^{(2)}_{\alpha\beta},$ $\dots\big\}$, and requiring terms at the $n$--th order to give, through the dynamics of $\Theta^{(n)}_{\alpha\beta}$, a conserved tensor field $\texttt{C}^{(n)}_{\alpha\beta}$. In (\ref{theta-expand}),  use has been made of $\Lambda \simeq (8 \pi G_N)^{-1}$ as follows from (\ref{newton}) thanks to the extreme smallness of $|\Lambda_0|$. Clearly, ${{\Theta}^{(1)}_{\alpha\beta}}$ in (\ref{theta-expand}) corresponds to $\Theta$ in (\ref{mat-T}), and ${{\Theta}^{(n\geq 2)}_{\alpha\beta}}$ represent the added features for achieving consistency in (\ref{eomppp}). 

Despite the structure (\ref{theta-expand}), $\texttt{C}^{(0)}_{\alpha\beta}$ and $\texttt{C}^{(1)}_{\alpha\beta}$ both stay put at their 
previous values in (\ref{C0value}) and (\ref{C1value}), respectively. The only difference is that $\Theta$ in (\ref{ek-cond}) is replaced by ${{\Theta}^{(1)}_{\alpha\beta}}$, and hence, what appears in  (\ref{soln-mathbbT}) are the first two terms of (\ref{theta-expand}). Consequently, at levels of  $n=0$ and $n=1$,  gravitational dynamics in (\ref{eomppp}) stay intact to the serial structure of $\mathbb{T}$ introduced in (\ref{theta-expand}). At the higher orders, $n\geq 2$, the situation changes due to the introduction of ${{\Theta}^{(n\geq 2)}_{\alpha\beta}}$.  For example, if $n=2$, the tensorial connection $\Delta^{\lambda}_{\alpha\beta}$ is quadratic in ${\Theta^{(1)}}_{\alpha\beta}$ and linear in ${\Theta^{(2)}}_{\alpha\beta}$
\begin{eqnarray}
\label{Delta-yeni}
\Delta^{\lambda}_{\alpha\beta}(2) =  8 \pi G_N \left( - {\Theta^{(1)}}^{\lambda}_{ \rho} \Delta^{\rho}_{\alpha\beta}\left(1\right)
+ 4 \pi G_N\, \left({\delta^{(2)}}\right)^{\lambda}_{\alpha\beta}\right)
\end{eqnarray}
which differs from (\ref{tensor-conn-n}) by the presence of
\begin{eqnarray}
\left({\delta^{(2)}}\right)^{\lambda}_{\alpha\beta} = \nabla_{\alpha} {\Theta^{(2)}}^{\lambda}_{\beta} + \nabla_{\beta} {\Theta^{(2)}}^{\lambda}_{\alpha} - \nabla^{\lambda} {\Theta^{(2)}}_{\alpha \beta} 
\end{eqnarray}
induced by ${\Theta^{(2)}}_{\alpha\beta}$ alone. Replacement of (\ref{Delta-yeni}) in  (\ref{eom}) yields ${\cal{O}}\left[\left(8\pi G_N\right)^2\right]$ terms which involve both ${\Theta^{(2)}}_{\alpha\beta}$ and ${\Theta^{(1)}}_{\alpha\beta}$, where the latter is related to $\texttt{t}_{\alpha\beta}$ via Eq.~(\ref{ek-cond}). 

The Bianchi-wise consistency and completeness of Einstein field equations are based on the feature that the three tensor fields, $G_{\alpha\beta}\left(\between_V, V\right)$,  $\texttt{C}^{(0)}_{\alpha\beta}$ and $\texttt{C}^{(1)}_{\alpha\beta}$, are the only divergence-free symmetric tensor fields in 4-dimensional spacetime \cite{lovelock}. There exist no other divergence-free, symmetric tensor fields with which $\texttt{C}^{(n\geq 2)}_{\alpha\beta}$ can be identified. In fact, there is no analogue of Huggins tensor in curved space \cite{huggins, lovelock}. Consequently, instead of strict vanishing of the divergences of $\texttt{C}^{(n\geq 2)}_{\alpha\beta}$,  which cannot be achieved,  one must be content with non-vanishing yet higher order remnants to be canceled by divergences of higher orders.  More accurately, if divergence of $\texttt{C}^{(n)}_{\alpha\beta}$, in the equation of motion (\ref{eompp}),  gives a remnant at order of $(n+1)$-st and higher  then divergence at the $n$-th level is effectively nullified.  At the $n=2$ level, for instance, one can consider the tensor field
\begin{eqnarray}
\label{c2example}
\texttt{C}^{(2)}_{\alpha\beta} &=& \Big(- \boxminus_{\alpha\beta} g_{\mu \nu} + \boxminus_{\alpha\mu} g_{\beta \nu} + \boxminus_{\beta\mu} g_{\alpha\nu} - \nabla_{\mu}\nabla_{\nu} g_{\alpha\beta} - 2 {G}_{\alpha\mu\beta\nu} \nonumber\\ &+&  \frac{1}{2} \Box \left( 2 g_{\mu \nu} g_{\alpha\beta} -   g_{\alpha\mu} g_{\beta\nu} - g_{\alpha\nu} g_{\beta\mu}\right) \Big) \Omega^{\mu\nu}
\end{eqnarray}
where $\boxminus_{\alpha\beta} \equiv \nabla_{\alpha}\nabla_{\beta} - G_{\alpha\beta}$, ${G}_{\alpha\mu\beta\nu} \equiv  R_{\alpha\mu\beta\nu} - \frac{1}{2} g_{\alpha\beta} R_{\mu\nu}$, and 
\begin{eqnarray}
\Omega_{\alpha\beta} = c_1 {\Theta^{(1)}}_{\alpha}^{\mu} {\Theta^{(1)}}_{\mu \beta} + c_2 {\Theta^{(1)}}^{\mu}_{\mu} {\Theta^{(1)}}_{\alpha\beta} + c_3 {\Theta^{(1)}}^{\mu}_{\mu} {\Theta^{(1)}}^{\nu}_{\nu} g_{\alpha\beta} + c_4 \texttt{t}_{\alpha\beta}
\end{eqnarray}
with $c_{1,\dots,4}$ being dimensionless constants. Obviously, divergence of $\Omega_{\alpha\beta}$ does not vanish, and it is non-local due to its dependence on ${\Theta^{(1)}}_{\alpha\beta}$. Expectedly, divergence of $\texttt{C}^{(2)}_{\alpha\beta}$ does not vanish yet it is ${\cal{O}}\left[(8 \pi G_N) \texttt{t}\nabla \Omega \right]$ on the equation of motion (\ref{eompp}). It is sufficiently suppressed since it falls at the $n=4$ order, and may be made to cancel with the divergence of $n=4$ term. This progressive, systematic cancellation works well as long as divergence of $\texttt{C}^{(n)}_{\alpha\beta}$ produces terms at the $n$--th and $(n+1)$--st orders so that
the $n$--th order term cancels the non-vanishing divergence coming from the $(n-1)$--st order. This procedure, order by order in $(8 \pi G_N)$, adjusts $\mathbb{T}_{\alpha\beta}$, more correctly its $\Theta_{\alpha\beta}$ part, to guarantee the conservation of matter and radiation flow.

In general, the mechanism proposed involves higher powers of $G_N$ associated with higher powers of  ${\Theta^{(n)}}$ encoding the matter sector. Accordingly, the dynamical equations are expected to involve higher powers of curvature tensors. These higher order contributions from either sector are constrained by the Bianchi identities. In fact, $\texttt{C}^{(n)}_{\alpha\beta}$ encode nothing but these mutual contributions from material and gravitational sectors. This is best illustrated by $\texttt{C}^{(2)}_{\alpha\beta}$ in (\ref{c2example}): Curvature tensors and covariant derivatives acting on $\Omega_{\alpha\beta}$ are collected together to make the divergence of $\texttt{C}^{(2)}_{\alpha\beta}$ higher order. 

Also, one notes that the expression of $\texttt{C}^{(2)}_{\alpha\beta}$ in (\ref{c2example}) serves only as an illustration. It is obviously not exhaustive, as $\texttt{C}^{(2)}_{\alpha\beta}$ cannot be guaranteed to depend on ${\Theta^{(1)}}$ through only $\Omega$. It may well involve structures like  $\nabla {\Theta^{(1)}} \nabla {\Theta^{(1)}}$ or $ {\Theta^{(1)}} \nabla\nabla {\Theta^{(1)}}$.  One also notes that, however it is composed of ${\Theta^{(1)}}_{\alpha\beta}$ and ${\texttt{t}}_{\alpha\beta}$,  $\Omega_{\alpha\beta}$ originates from ${\Theta^{(2)}}_{\alpha\beta}$ as the remnant of competing ${\Theta^{(1)}}$--   and ${\Theta^{(2)}}$--dependent parts of (\ref{Delta-yeni}). Essentially, what is happening is that  ${\Theta^{(2)}}_{\alpha\beta}$ gets expressed in terms of ${\Theta^{(1)}}_{\alpha\beta}$ via $\Omega_{\alpha\beta}$ so that the divergence of $\texttt{C}^{(2)}_{\alpha\beta}$ jumps to $n=4$ level.

\subsection{ Answer to Question 3}
Having arrived at the gravitational field equations (\ref{eomppp}), it is clear that $\Lambda_0$ stands out as the only dark energy source to account for the observational value of the CC \cite{astro8}. In other words, one is left with the identification
\begin{eqnarray}
\label{cond-x}
\Lambda_{\texttt{eff}} = \Lambda_0  \lesssim H_0^2
\end{eqnarray}
to be constrasted with (\ref{lameff}) in GR. It is manifest that this result involves no fine or coarse tuning  of distinct curvature sources. The vacuum energy $\texttt{E}$, instead of gravitating, generates the gravitational constant $G_N$ via
\begin{eqnarray}
\label{xcxcx}
\left(8 \pi G_N\right)^{-1} \simeq \texttt{L}^2 \texttt{E} 
\end{eqnarray}
where $\texttt{L}^2$ is an area parameter which converts the vacuum energy into Newton's constant. This parameter is not fixed by the model. Essentially, it adjusts itself against possible variations in vacuum energy density $\texttt{E}$ so that $G_N$ is correctly generated. If $\texttt{E} \sim \left(M_{EW}\right)^4$ then $\texttt{L}^2 \sim m_{\nu}^{-2}$. In this scenario, contributions to vacuum energy from   quantal matter whose loops smaller than the electroweak scale are canceled by some symmetry principle. Low-energy supersymmetry is  this sort of symmetry. On the other hand, if $\texttt{E} \sim \left(8 \pi G_N\right)^{-2}$ then $\texttt{L}\sim \ell_{Pl}$. In this case vacuum energy stays uncut up to the Planck scale, and $\texttt{E}$ and $\texttt{L}^2$ happen to be determined by a single scale. Therefore, this case turns out to be the most natural one compared to cases where the vacuum energy falls to an intermediate scale.  In a sense, the worst case of GR translates into the best case of the present scenario. 

As was also noted in \cite{demirz}, the result (\ref{xcxcx}) guarantees that matter and radiation are prohibited from causing the CCP. In spite of this, one must keep in mind that quantum gravitational effects can restore the CCP by shifting $\Lambda_0$ by quartically-divergent contribution of the graviton and graviton-matter loops. If gravity is classical, however, the mechanism successfully avoids the CCP by canalizing the vacuum energy deposited by quantal matter into the generation of the gravitational constant.   Namely, stress-energy connection alters the role and meaning of the vacuum energy in a striking way.  Newton's constant is the outlet of the vacuum energy.  

A critical aspect of the mechanism, which has  not been mentioned so far, is that the seed dynamical equations (\ref{eom-new}) do not follow from an action principle. Indeed,  the germ of the mechanism rests entirely upon the matter-free gravitational field equations in GR, and it is not obvious if it can ever follow from an action principle. Though one can argue for the Einstein-Hilbert action at the linear level in (\ref{eomppp}), the non-local, higher-order terms do not fit into this picture. Thus, one concludes that, gravitational field equations at finis involve non-local, Planck-suppressed higher-order effects, and they are difficult, if not impossible, to derive from an action principle.  

\section{Conclusion}
The CCP is too perplexing to admit a resolution within the GR or quantum field theory. Any attempt at adjudicating the problem is immediately faced with the conundrum that the fundamental equations are to be processed to offer a resolution for the CCP by maintaining all the successes of quantum field theory and GR.

In the present work, gravity is taken classical yet matter and radiation are interpreted as quantal. The vacuum energy deposited by quantal matter and its gravitational consequences are explored in complete generality by erecting a non-Riemannian geometry on the stress-energy tensor. By using the scaling properties of gravitational field equations in GR as a giude, it has been inferred that stress-energy tensor can  be incorporated into gravitational dynamics by modifying the connection. This observation gives rise to a novel framework in which the gravitational constant $G_N$ derives from the vacuum energy. In fact, vacuum energy, instead of curving the spacetime, happens to generate the gravitational constant. Indeed, contrary to GR, the vacuum energy induced by quantal matter is not `cosmological constant'; it just sources the `gravitational constant'.  The CC stays put at its bare value, and its identification with the observational value involves no tuning of distinct quantities as long as gravity is classical. Quantum gravitational effects bring back the CCP by adding to $\Lambda_0$ quartically-divergent  contributions of the graviton and graviton-matter loops. 

In spite of these observations, the model is in want of certain rectifications for a number of vague aspects. One of them is the absence of an action principle.  Another aspect concerns a complete analysis of the quantum gravitational effects. Another point to note is the parameter $\texttt{L}^2$ whose dynamical origin is obscure. Finally, the case $|\Theta| \lesssim |\Lambda|$ must be studied in depth to determine strong gravitational effects. All these points and many not mentioned here are topics of further analyses of the model. 

The literature consists of numerous attempts at solving the CCP. The proposals conceptually and practically vary in a rather wide range (See the long list of references in the review volumes \cite{review,weinberg1,nobbenhuis1} and in \cite{demirz}.  Recent work based on extended gravity theoeries are given in \cite{recent}). The mechanism proposed in this work, which significantly improves and expands \cite{demirz},  differs from those in the literature by its ability to tame the vacuum energy induced by already known physics down to the terscale, by its immunity to any symmetry principle beyond general covariance, and by  its originatility in canalizing the vacuum energy to generation of the gravitational constant. 
 
\section{Acknowledgements} 
The author is grateful to N. Dadhich,  C. Germani, V. Husain and S. Randjbar-Daemi for fruitful conversations and discussions. He thanks Anne Frary for her editing of the manuscript. Also, he is thankful to the anonymous referee for his/her critical and constructive comments.

\end{document}